\documentclass[11pt]{article}

\usepackage[preprint]{acl}

\usepackage{times}
\usepackage{latexsym}

\usepackage[T1]{fontenc}
\usepackage[utf8]{inputenc}
\usepackage{microtype}
\usepackage{inconsolata}

\usepackage{graphicx}
\usepackage{enumitem}
\usepackage{tikz}
\usepackage{pgfplots}
\usepackage{url}
\usepackage{array}
\usepackage{tabularx}
\usepackage{booktabs}
\usetikzlibrary{arrows.meta, positioning, shapes.geometric, fit, backgrounds}
\pgfplotsset{compat=1.18}
\title{Toward Human-Centered Multi-Agent Systems:\\Integrating Cognition, Culture, Values, and Cooperation in AI Agents}

\author{Safia Baloch \\
  GIK Institute \\
  Topi, KPK, Pakistan \\
  \texttt{safia.baloch@giki.edu.pk} \\\And
  Rahemeen Khan \\
  University of Milano-Bicocca \\
  Milano, Itlay \\
  \texttt{rahemeen\_ahmed@yahoo.com}}

\begin{document}
\maketitle

\begin{abstract}
The emergence of large language model (LLM)-based agents and multi-agent systems has accelerated a transition in artificial intelligence from narrow task automation to increasingly autonomous decision-making. Yet, despite remarkable progress in language generation, planning, tool use, and coordination, most contemporary agents still treat intelligence primarily as prediction, optimization, and task completion. Human environments, however, are fundamentally social and normative: people reason under bounded rationality, communicate through culturally situated language, make decisions through values and beliefs, and collaborate through trust, shared mental models, and social norms. This survey argues that the next generation of AI agents especially those acting on behalf of humans must move beyond task competence toward \emph{human-centered} capabilities.

We synthesize research across six interconnected areas: (1) the evolution of intelligent agents, (2) human cognition and decision-making, (3) language, culture, and social context, (4) human values and belief systems, (5) human-agent collaboration, and (6) multi-agent coordination and computational modeling of human characteristics. We review foundational theories from cognitive science, sociolinguistics, computational social science, and AI alignment, alongside recent work on LLM agents, cultural alignment benchmarks, preference learning, explainability, and agent societies. Based on this synthesis, we identify a central research gap: while substantial progress has been made in each dimension individually, there remains no widely adopted framework that jointly integrates cognition, culture, values, and social behavior into autonomous agents capable of authentic human-centered collaboration. We conclude by outlining future research directions for building culturally aware, value-aligned, cognitively grounded, and cooperative multi-agent systems.
\end{abstract}

\section{Introduction}

Artificial intelligence has entered an era in which agents are no longer limited to narrow task execution. The rise of foundation models and LLMs has enabled systems that can perceive instructions, reason over long contexts, call tools, simulate social behavior, and coordinate with other agents \citep{brown2020language, wang2023autonomous, guo2024llmmas, chen2025llmmas}. This shift has fueled an emerging paradigm sometimes described as \emph{agentic AI}: systems capable not merely of responding, but of acting.

At the same time, a conceptual limitation has become increasingly visible. Much of current AI still models intelligence as a combination of prediction, optimization, and procedural task completion \citep{russell2019human, ji2024alignment}. In contrast, human intelligence is inseparable from social context. Human reasoning is often bounded rather than optimal \citep{simon1957models}; judgments are shaped by heuristics and biases \citep{tversky1974judgment, kahneman2011thinking}; communication depends on pragmatics, social identity, and culture \citep{hall1976beyond, hofstede2001culture}; and decisions are influenced not only by goals but by values, norms, and beliefs \citep{russell2019human, ji2024alignment}. Humans also collaborate through trust, role awareness, negotiation, and shared mental models \citep{salas2005shared, doshivelez2017rigorous}.

These observations matter because AI agents are increasingly deployed in roles traditionally occupied by humans: assistants, advisors, mediators, teammates, and autonomous participants in organizational workflows. In such settings, success cannot be measured only by accuracy or efficiency. Agents acting on behalf of humans must be able to preserve and reflect human-centered properties: culturally appropriate language, context-sensitive reasoning, value-aligned decision making, and socially adaptive collaboration.

This survey develops the argument that future multi-agent systems should not merely be \emph{intelligent}; they should be \emph{human-centered}. We define human-centered multi-agent systems as systems in which agents are designed not only to solve tasks, but also to reason and interact in ways that account for human cognition, culture, values, and cooperative dynamics.

The survey is organized around the following progression:

\begin{itemize}[leftmargin=*]
    \item We first review the \textbf{evolution of intelligent agents}, from rule-based systems to LLM-based autonomous and multi-agent architectures.
    \item We then examine \textbf{human cognition and decision-making}, focusing on bounded rationality, dual-process reasoning, common sense, and theory of mind.
    \item Next, we discuss \textbf{language, culture, and social context}, including cross-cultural psychology, sociolinguistics, and culturally adaptive language technologies.
    \item We then consider \textbf{human values and belief systems}, drawing on AI alignment, preference learning, and machine ethics.
    \item We connect these attributes to \textbf{human-agent collaboration}, emphasizing trust, explainability, and shared mental models.
    \item Finally, we analyze \textbf{multi-agent systems and computational modeling of human characteristics}, highlighting the gap between coordination-centric architectures and richer human-centered representations.
\end{itemize}

The central claim of this paper is straightforward: \textbf{existing AI and multi-agent systems primarily focus on intelligence as prediction, optimization, and task completion, whereas agents acting on behalf of humans in cooperative environments require deeper computational representations of human cognition, culture, values, and social behavior.} While significant progress has been made in each of these topics independently, the literature still lacks unified frameworks that integrate them into robust autonomous agents.

\section{A Unified View of Human-Centered Agents}

\begin{figure*}[t]
\centering
\begin{tikzpicture}[
    node distance=0.9cm and 1.0cm,
    box/.style={draw, rounded corners=8pt, thick, align=center, minimum width=4.1cm, minimum height=1.2cm, font=\small},
    smallbox/.style={draw, rounded corners=8pt, thick, align=left, minimum width=3.9cm, minimum height=1.0cm, font=\small},
    pillar/.style={draw, rounded corners=10pt, thick, minimum width=3.3cm, minimum height=5.0cm, align=center, font=\small},
    base/.style={draw, rounded corners=10pt, thick, minimum width=14.2cm, minimum height=1.2cm, align=center, font=\small},
    title/.style={font=\bfseries\small}
]

\node[base, fill=gray!10] (base) {Multi-Agent System Infrastructure\\
\textit{communication protocols, memory, planning, coordination, tool use, distributed decision-making}};

\node[pillar, fill=blue!7, above left=1.1cm and -0.3cm of base.north west, anchor=south west] (cog) {\textbf{Cognition}\\[0.2cm]
\begin{tabular}{l}
$\bullet$ bounded rationality\\
$\bullet$ dual-process reasoning\\
$\bullet$ common-sense reasoning\\
$\bullet$ mental models\\
$\bullet$ theory of mind
\end{tabular}};

\node[pillar, fill=green!7, right=0.7cm of cog] (cult) {\textbf{Culture \& Social Context}\\[0.2cm]
\begin{tabular}{l}
$\bullet$ sociolinguistics\\
$\bullet$ cultural norms\\
$\bullet$ contextual communication\\
$\bullet$ regional variation\\
$\bullet$ social identity
\end{tabular}};

\node[pillar, fill=orange!8, right=0.7cm of cult] (vals) {\textbf{Values \& Beliefs}\\[0.2cm]
\begin{tabular}{l}
$\bullet$ value alignment\\
$\bullet$ preference learning\\
$\bullet$ pluralistic values\\
$\bullet$ moral reasoning\\
$\bullet$ belief representation
\end{tabular}};

\node[pillar, fill=purple!8, right=0.7cm of vals] (coop) {\textbf{Cooperation}\\[0.2cm]
\begin{tabular}{l}
$\bullet$ trust\\
$\bullet$ explainability\\
$\bullet$ shared mental models\\
$\bullet$ teamwork\\
$\bullet$ negotiation
\end{tabular}};

\node[box, fill=yellow!12, above=0.7cm of cult.north, minimum width=7.8cm] (goal) {\textbf{Human-Centered Agent Behavior}\\
context-aware, culturally adaptive, value-aligned, and socially cooperative decision-making};

\draw[-{Latex[length=3mm]}, thick] (base.north) -- (cog.south);
\draw[-{Latex[length=3mm]}, thick] (base.north) -- (cult.south);
\draw[-{Latex[length=3mm]}, thick] (base.north) -- (vals.south);
\draw[-{Latex[length=3mm]}, thick] (base.north) -- (coop.south);

\draw[-{Latex[length=3mm]}, thick] (cog.north) -- (goal.south west);
\draw[-{Latex[length=3mm]}, thick] (cult.north) -- (goal.south);
\draw[-{Latex[length=3mm]}, thick] (vals.north) -- (goal.south east);
\draw[-{Latex[length=3mm]}, thick] (coop.north) to[out=90,in=330] (goal.east);

\end{tikzpicture}
\caption{A unified conceptual framework for human-centered multi-agent systems. Contemporary agents typically rely on a multi-agent infrastructure for planning, communication, and task execution. This survey argues that agents acting on behalf of humans must additionally model cognition, culture, values, and cooperation in an integrated manner.}
\label{fig:framework}
\end{figure*}
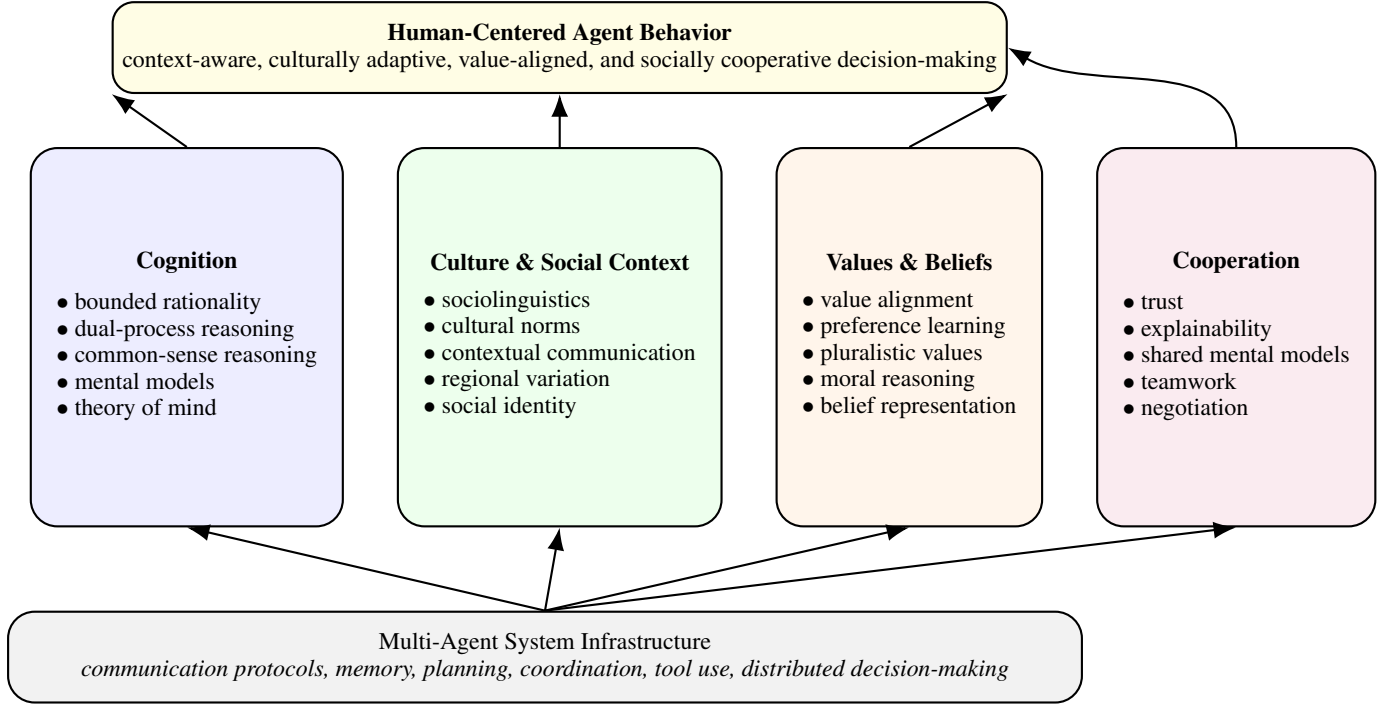

Before reviewing the literature in detail, it is useful to articulate the conceptual space this survey addresses. Figure~\ref{fig:framework} presents a unified perspective on human-centered agents. The framework places a multi-agent infrastructure at the base, but argues that effective participation in socially embedded environments requires four additional layers: cognition, culture, values, and cooperation.

The figure is intentionally layered rather than modular. It emphasizes that human-centered behavior is not a single capability added to a task-oriented agent; it emerges from the interaction of multiple dimensions. Cognition without culture yields abstract reasoning detached from context. Culture without values risks superficial adaptation. Values without cooperation produce rigid agents that may be aligned in isolation but ineffective in teams. The broader research challenge, therefore, is integration.


\section{Evolution of Intelligent Agents}

\subsection{From Rule-Based Automation to Learning Systems}

The history of intelligent agents can be read as a sequence of increasing representational and decision-making flexibility. Early AI systems were primarily rule-based. Classical expert systems and symbolic agents relied on hand-crafted knowledge bases and inference rules to solve well-defined problems \citep{nilsson1980principles, russellnorvig2021}. These systems excelled in constrained domains but suffered from brittleness, poor scalability, and limited adaptability.

Machine learning shifted the field away from manual rule-writing toward data-driven generalization \citep{bishop2006pattern}. Statistical NLP, reinforcement learning, and deep learning substantially improved performance across perception, classification, language understanding, and control. Yet most such systems still optimized narrowly defined objectives, often without explicit representations of social context or human preferences.

\subsection{Foundation Models and the Re-emergence of Agents}

The introduction of large-scale pretrained language models marked a major reconfiguration of the agent discussion. Models such as GPT-3 demonstrated remarkable in-context learning and language generation abilities \citep{brown2020language}. Subsequent work revealed that LLMs can support planning, reflection, tool use, role specialization, and long-horizon workflows, leading to a rapid expansion of research on autonomous and multi-agent LLM systems \citep{wang2023autonomous, xi2023rise, guo2024llmmas, chen2025llmmas}.

Recent surveys have converged on the idea that LLM-based agents can be decomposed into components such as profiling, memory, planning, action, and feedback \citep{wang2023autonomous, xi2023rise}. In multi-agent settings, research has explored role assignment, collaboration structures, communication protocols, and collective problem solving \citep{guo2024llmmas, tran2025multiagent}. These architectures expand the operational scope of AI from single-turn response generation to persistent action in complex environments.

\subsection{The Remaining Limitation}

Despite these advances, the dominant design logic of many current agents remains task-centric. Agents may communicate fluently, invoke tools effectively, and coordinate procedurally, yet still fail to capture key dimensions of human behavior. They often exhibit weak grounding in social norms, inconsistent persona maintenance, limited cultural adaptation, and narrow notions of alignment \citep{cao2023crosscultural, alkhamissi2024cultural, wang2024cdeval, baltaji2024persona}. In other words, modern agents are increasingly autonomous, but not yet deeply human-centered.

\subsection{Research Gap}

The evolution of intelligent agents reveals a widening gap between \emph{capability} and \emph{human representation}. Contemporary agents have gained powerful mechanisms for task decomposition, communication, and action, but they still lack unified representations of how humans reason, what humans value, and how humans behave across cultural and social settings. This gap motivates the rest of the survey.

\section{Human Cognition and Decision-Making}

If agents are expected to act on behalf of humans, they must do more than optimize utility functions or rank likely next actions. They must approximate how humans reason under uncertainty, constraints, and social context.

\subsection{Rationality and Its Limits}

Classical decision theory often assumes rational actors who optimize expected utility. While such models remain foundational, decades of cognitive research show that human reasoning deviates systematically from strict rationality. \citet{simon1957models} introduced \emph{bounded rationality} to describe how humans make satisfactory rather than optimal decisions under limited time, information, and cognitive resources. This insight is highly relevant for agent design: human-centered agents should not be modeled as perfect optimizers detached from realistic decision conditions.

\citet{tversky1974judgment} and \citet{kahneman2011thinking} further demonstrated that human judgment is shaped by heuristics and biases. People rely on mental shortcuts such as availability and representativeness, which can be efficient but also error-prone. The implication is not that AI should reproduce human biases uncritically, but that human decision-making is not reducible to clean optimization. Agents intended to support or imitate human reasoning must account for this hybrid structure of efficiency, boundedness, and context dependence.

\subsection{Dual-Process Accounts of Reasoning}

Dual-process theories distinguish between fast, intuitive reasoning (often called System~1) and slow, deliberative reasoning (System~2) \citep{kahneman2011thinking, evans2008dual}. This distinction has become increasingly relevant in AI. Many LLM systems appear strong at rapid pattern completion and associative inference, yet struggle with deliberate, grounded, or multi-step reasoning unless scaffolded by prompting, decomposition, or external tools \citep{wei2022cot, kojima2022zeroshot, huang2023reasoning}.

This parallel has motivated interest in cognitive hybrids that combine neural language models with more structured reasoning or cognitive architectures. For example, recent work explores integrating ACT-R-inspired mechanisms into LLM systems to ground perception, memory, and deliberate decision-making \citep{wu2024cognitivellm}. Such work reflects a broader recognition that human-like reasoning may require architectures capable of shifting between intuitive and reflective modes.

\subsection{Common-Sense Reasoning}

Human reasoning is deeply supported by common sense: tacit world knowledge about causality, temporality, physical interaction, social routines, and practical inference. Common-sense reasoning has long been recognized as a core AI challenge \citep{mccarthy1959programs}. While LLMs often display impressive common-sense behavior in language tasks, their competence remains uneven and benchmark-dependent \citep{huang2023reasoning}. In human-centered agent settings, common sense is not merely a benchmark ability; it is a requirement for believable, safe, and contextually appropriate action.

\subsection{Theory of Mind and Mental Models}

A particularly important aspect of social cognition is \emph{Theory of Mind} (ToM): the ability to reason about the beliefs, desires, intentions, and knowledge states of others \citep{premack1978does}. ToM underlies persuasion, explanation, coordination, and conflict resolution. In collaborative settings, humans continuously model what others know, what they expect, and how they are likely to react.

Recent work has begun evaluating ToM-like capabilities in LLMs and surveying related benchmarks \citep{nguyen2025tomsurvey, chen2025tom}. Findings suggest that LLMs can perform well on some stylized tasks, yet remain non-robust and sensitive to prompt design, dataset artifacts, and superficial cues. This distinction is crucial: high benchmark performance does not necessarily imply stable, deployable social reasoning. For multi-agent and human-facing systems, ToM must be reliable, contextual, and interaction-aware.

Relatedly, the literature on \emph{mental models} in teamwork emphasizes that effective collaboration relies on partially shared internal representations of goals, roles, constraints, and plans \citep{salas2005shared}. Human-centered agents should not only infer others' mental states, but also maintain and update interoperable representations of what is jointly understood.

\subsection{Implications for Agent Design}

Taken together, the literature suggests that representing human reasoning in AI requires at least five ingredients:
\begin{enumerate}[leftmargin=*]
    \item bounded rather than perfectly rational decision processes,
    \item the ability to combine fast heuristics with slow deliberation,
    \item common-sense world knowledge,
    \item ToM-like reasoning about others' beliefs and intentions, and
    \item dynamic mental models of shared tasks and social interactions.
\end{enumerate}

Current agents often implement fragments of this picture, but rarely the whole. The result is a recurring mismatch: agents can complete tasks but still behave in ways that feel inhuman, socially clumsy, or insensitive to context.

\section{Language, Culture, and Social Context}

One of the most distinctive and underdeveloped aspects of human-centered agents is cultural and sociolinguistic adaptation. Human communication is never purely semantic. Meaning is mediated by culture, audience, norms, and interactional expectations.

\subsection{Culture as a Determinant of Behavior}

Cross-cultural psychology provides foundational evidence that values, communication styles, and behavioral expectations vary systematically across societies. \citet{hofstede2001culture} characterized differences along dimensions such as individualism vs. collectivism, power distance, uncertainty avoidance, masculinity/femininity, long-term orientation, and indulgence/restraint. \citet{hall1976beyond} further distinguished between high-context and low-context communication, highlighting how different cultures encode meaning explicitly or implicitly.

These frameworks are, of course, abstractions and have been critiqued when used simplistically. Nevertheless, they remain influential reference points for understanding cultural variability. For AI, the key lesson is that the ``same'' instruction, explanation, or negotiation strategy may be interpreted differently across cultural settings.

\subsection{Sociolinguistics and Pragmatics}

Sociolinguistics shows that language varies as a function of region, class, gender, institutional context, and community norms. Meaning arises not only from literal content but from style, stance, presupposition, and politeness. As a result, agents that are fluent but socially generic may still be culturally inappropriate or pragmatically ineffective.

This issue becomes especially important in collaborative environments where language performs social work: signaling respect, maintaining face, negotiating disagreement, or expressing uncertainty. Human-centered agents must therefore go beyond grammatical fluency toward pragmatic and sociocultural competence.

\subsection{Cultural Alignment in LLMs}

Recent NLP research has begun to investigate whether LLMs reflect cultural diversity or disproportionately encode dominant linguistic and ideological patterns. \citet{cao2023crosscultural} show that ChatGPT aligns more strongly with American cultural patterns and adapts less effectively to other contexts, while English prompting can flatten cross-cultural differences. \citet{alkhamissi2024cultural} find that cultural alignment improves when models are prompted in the dominant language of a culture and when pretraining better reflects relevant language mixtures. \citet{wang2024cdeval} introduce CDEval, a benchmark measuring cultural dimensions in LLM behavior, and show that models vary substantially across domains and dimensions. \citet{masoud2025hofstede} similarly evaluate cultural alignment through Hofstede-inspired analysis and report uneven adaptation across regions and languages.

Together, these studies establish an important point: current models can appear globally deployable while remaining culturally asymmetric. They often reflect a narrow center of gravity rather than genuinely pluralistic social understanding.

\subsection{Context-Aware Dialogue and Social Intelligence}

Another line of work examines context-aware and socially aware dialogue systems. While recent models are far more fluent than earlier dialogue systems, contextual adaptation often remains shallow: models can maintain local coherence but may miss deeper social cues, interactional histories, or culturally embedded expectations. This gap is amplified in high-stakes settings such as healthcare, education, administration, and conflict mediation, where how something is said may be as important as what is said.

\subsection{Implications for Human-Centered Agents}

Culturally and socially aware agents require more than multilingual capacity. They need:
\begin{itemize}[leftmargin=*]
    \item representations of social and cultural context,
    \item adaptive communication strategies sensitive to audience norms,
    \item mechanisms to avoid collapsing diverse value systems into a single default style, and
    \item evaluation frameworks that measure not only correctness, but cultural appropriateness and social fit.
\end{itemize}

This is one of the clearest gaps in current agent research: language generation has advanced rapidly, but genuine cultural understanding remains limited.

\section{Human Values and Belief Systems}

An agent acting on behalf of humans must not only understand what people \emph{say}; it must also reason about what they \emph{care about}. This brings the survey into the domain of alignment, ethics, preference learning, and computational representations of belief.

\subsection{From Objective Optimization to Alignment}

As AI systems become more capable, alignment has emerged as a central research concern. Broadly speaking, alignment asks whether system behavior accords with human intentions and values \citep{russell2019human, ji2024alignment}. This issue becomes particularly important when agents make autonomous decisions under uncertainty or operate over extended horizons.

A key challenge is that many AI systems optimize measurable proxies rather than human goals themselves. Reward misspecification, distribution shift, and objective gaming can all produce behavior that is technically competent but normatively undesirable \citep{amodei2016problems, ji2024alignment}. In multi-agent contexts, this challenge becomes dynamic: interaction among agents may amplify or distort misalignment \citep{carichon2025multiagent}.

\subsection{Preference Learning}

Preference learning has become one of the most influential approaches to alignment in foundation models. Early work on inverse reinforcement learning sought to infer reward functions from observed behavior \citep{ng2000algorithms}. More recent approaches include reinforcement learning from human feedback (RLHF) \citep{ouyang2022training}, direct preference optimization (DPO), constitutional methods, and broader forms of human preference modeling.

Recent surveys provide a detailed taxonomy of how preferences are sourced, represented, and optimized in LLM alignment \citep{jiang2024preference, gao2024unifiedpreference}. These works make clear that preference alignment is not a single technique but an ecosystem of choices involving feedback collection, reward modeling, policy optimization, and evaluation.

\subsection{Pluralistic and Personalized Alignment}

A particularly important development for your topic is the move from generic ``human preference'' to \emph{pluralistic} and \emph{personalized} alignment. Human values are not homogeneous. They vary across individuals, communities, and cultures. Recent surveys on pluralistic and personalized alignment argue that one-size-fits-all alignment is insufficient for real-world deployment \citep{xie2025pluralistic, guan2025personalized}. Benchmarks such as PERSONA emphasize the challenge of representing diverse and potentially conflicting user value profiles \citep{castricato2025persona}.

\subsection{Belief Representation and Moral Reasoning}

Modeling values also requires modeling beliefs. Human behavior depends on what people perceive to be true, what they regard as important, and which norms they think apply in a given context. Some recent work proposes value--belief--norm reasoning as a means of improving opinion alignment and persona-sensitive prediction \citep{do2025coo}. This is promising because it moves beyond static persona labels toward explanatory structures linking values, beliefs, and decisions.

Computational ethics and machine ethics further ask how moral concerns can be represented algorithmically \citep{wallach2009moral}. Although the field remains philosophically and technically unresolved, its relevance for agent systems is undeniable. Agents deployed in socially embedded roles must make choices that affect autonomy, fairness, confidentiality, harm, respect, and legitimacy.

\subsection{Implications for Human-Centered Agents}

The literature suggests that value-aware agents need:
\begin{enumerate}[leftmargin=*]
    \item methods for learning and updating human preferences,
    \item support for pluralism rather than a single universalized user model,
    \item explicit mechanisms for linking beliefs, values, and decisions,
    \item safeguards against misalignment in interactive and multi-agent settings, and
    \item evaluation standards that capture normative adequacy rather than only task success.
\end{enumerate}

This remains one of the largest unresolved areas in human-centered agent design. Current systems often optimize a measurable objective, but real-world human representation demands richer normative modeling.

\section{Human-Agent Collaboration}

If agents are to become teammates rather than tools, they must collaborate in ways that humans perceive as intelligible, trustworthy, and adaptive. Human-agent collaboration is therefore not a peripheral application area; it is a test of whether human-centered design has succeeded.

\subsection{From Automation to Teaming}

Traditional AI systems often operated as automation tools: they executed subtasks delegated by humans. Human-agent collaboration moves beyond this paradigm toward \emph{joint activity}, where humans and agents contribute complementary strengths. Recent overviews from NLP and HCI emphasize that the key research question is no longer whether AI can perform tasks alone, but how AI and humans can work together most effectively \citep{wu2025humanai, yangwu2024hai}.

This reframing matters because collaboration requires more than capability. A strong autonomous solver may still be a poor teammate if it fails to communicate uncertainty, infer preferences, or coordinate interactively.

\subsection{Trust and Explainability}

Trust is central to collaboration. People are more likely to rely on systems that are transparent, predictable, and responsive. Explainable AI has long been motivated by the need to render system behavior understandable \citep{doshivelez2017rigorous}. In the LLM era, this concern has expanded into a rapidly growing research area on explainable and transparent language models \citep{cambria2024xai, palikhe2025transparent}.

However, explainability should not be reduced to post-hoc verbalization. In collaborative settings, useful explanations must be relevant to the user's goals, calibrated to expertise, and actionable. This is especially important in high-stakes domains where over-trust and under-trust can both be harmful.

\subsection{Shared Mental Models and Mixed-Initiative Interaction}

The teamwork literature stresses the importance of shared mental models: collaborators need an overlapping understanding of task structure, roles, and expectations \citep{salas2005shared}. In human-agent interaction, this implies that agents should be capable of communicating plans, updating goals, and negotiating task boundaries.

Recent work on decision-oriented dialogue formalizes scenarios in which AI assistants collaborate with humans to reach complex decisions, showing that current language models still underperform human assistants despite fluent dialogue \citep{lin2024decision}. Similarly, work on reinforcement learning-based human-agent collaboration for complex tasks shows that strategic human intervention can substantially improve outcomes when agents are not fully reliable \citep{feng2024rehac}. These studies collectively suggest that collaboration should be treated as an optimization target in its own right, not merely as a side-effect of agent competence.

\subsection{What Makes an Agent a Teammate?}

An effective teammate typically exhibits:
\begin{itemize}[leftmargin=*]
    \item awareness of the human's goals, limitations, and preferences,
    \item timely and context-appropriate communication,
    \item transparency about confidence and rationale,
    \item adaptive turn-taking and initiative management,
    \item willingness to defer, ask, or clarify when necessary.
\end{itemize}

Most current agents only partially satisfy these criteria. They are often impressive assistants but inconsistent teammates.

\section{Multi-Agent Systems and Social Coordination}

Human-centeredness becomes even more challenging when moving from individual agents to \emph{societies of agents}. Multi-agent systems (MAS) have a long history in AI, but LLM-based multi-agent systems have revitalized the field by enabling natural-language communication, role specialization, and emergent collaboration \citep{wooldridge2009introduction, guo2024llmmas, chen2025llmmas, tran2025multiagent}.

\subsection{Classical Foundations}

Classical MAS research studied distributed problem solving, coordination, communication protocols, negotiation, and collective behavior \citep{wooldridge2009introduction}. Many applications involved autonomous entities with local information that needed to coordinate under partial observability or distributed objectives.

These foundations remain highly relevant. Modern LLM-based MAS often reinvent classical MAS patterns centralized orchestration, peer-to-peer communication, role differentiation, and iterative deliberation under a new language-based interface.

\subsection{LLM-Based Multi-Agent Systems}

Recent surveys show that LLM-based MAS are being applied to software development, planning, simulation, social science experiments, information seeking, and collaborative reasoning \citep{guo2024llmmas, chen2025llmmas, tran2025multiagent}. Their appeal lies in combining the flexibility of language-mediated interaction with the modularity of distributed agents.
The following systems reveal several problems related to research agenda. 
\begin{itemize}[leftmargin=*]
    \item persona drift and instability in multi-agent discussions,
    \item conformity and peer-pressure-like effects,
    \item weak preservation of cultural roles or identities,
    \item insufficient representation of value conflicts, and
    \item limited models of trust and social accountability.
\end{itemize}

For instance, recent work on persona inconstancy in multi-agent collaboration finds that even when agents are instructed to maintain cultural positions, they may conform or shift inconsistently during discussion \citep{baltaji2024persona}. This suggests that current multi-agent coordination mechanisms are socially thin: they coordinate discourse, but not necessarily identity, value, or norm-consistent behavior.

\subsection{The Problem of Multi-Agent Misalignment}

Another emerging theme is that alignment in MAS cannot be treated as a static property of isolated agents. \citet{carichon2025multiagent} argue that alignment in multi-agent systems is dynamic and interaction-dependent: even individually aligned agents may become misaligned through social organization, competition, or emergent incentives. 

\subsection{Toward Socially Grounded Coordination}

Culturally and cognitively aware agents require coordination mechanisms that go beyond message passing. They need ways to represent:
\begin{enumerate}[leftmargin=*]
    \item socially meaningful roles,
    \item culturally appropriate communication norms,
    \item trust relationships,
    \item collective and individual value constraints, and
    \item dynamic negotiation over conflicting preferences.
\end{enumerate}

This is where many current systems remain underdeveloped. They coordinate effectively at the task level, but not yet at the human-social level.

\section{Computational Modeling of Human Characteristics}

The themes reviewed so far are conceptually rich, but they must eventually be implemented. This raises a core engineering question: how can human cognition, social behavior, and value systems be represented computationally?

\subsection{Cognitive Architectures}

Cognitive architectures such as ACT-R and Soar were developed to model human reasoning, memory, and problem solving at a computational level \citep{anderson1996actr, newell1990unified, lebiere2014actr}. Their significance for present-day agent research lies in their explicitness. Unlike purely statistical models, cognitive architectures provide interpretable mechanisms for memory retrieval, production rules, goal management, and procedural control.

Although they predate modern LLMs, these architectures are increasingly relevant in neuro-symbolic and hybrid agent design. Integrating them with LLMs offers one path toward modeling slower, more structured cognition on top of fluent language generation \citep{wu2024cognitivellm}.

\subsection{Agent-Based Modeling and Social Simulation}

Agent-based modeling (ABM) provides another important implementation paradigm. Instead of modeling cognition in isolation, ABM simulates populations of interacting agents whose local rules generate emergent macro-level behavior \citep{bonabeau2002abm}. This tradition has long been used in economics, epidemiology, political science, and computational social science.

The recent rise of LLM-driven generative agents has renewed interest in the use of language models for simulating believable human-like social behavior \citep{park2023generative, wang2023humanoid}. Such systems show how memory, reflection, planning, and interaction can produce emergent social phenomena. However, they also reveal a key limitation: believable behavior does not necessarily imply faithful modeling of culture, values, or psychological diversity.

\subsection{Persona and User Modeling}

User and persona modeling aim to represent stable or evolving characteristics of individuals, such as preferences, demographics, habits, beliefs, or communication styles. Recent work has explored trainable role-playing agents \citep{shao2023characterllm}, persona prompting and its limitations \citep{hu2024personaeffect, giorgi2024subjectivity, lutz2025promptpersona}, dynamic persona updating \citep{chen2025deeper}, and recommendation-oriented persona modeling \citep{shi2025personax}.

 However, existing approaches often remain shallow or fragile. Persona labels can produce stereotyped outputs, and prompt-based personas may be inconsistent across tasks or interaction histories. Richer user modeling must therefore move toward interaction-grounded, belief- and value-aware representations rather than static descriptors alone.

\subsection{Digital Humans and Human-Like Agents}

Several recent systems explicitly aim to build human-like generative agents or digital characters \citep{park2023generative, wang2023humanoid, shao2023characterllm}. These efforts are important because they make the problem concrete: what kinds of memory, reflection, emotion, need states, and persona structures make an artificial agent behave more believably like a person?

Yet believability and \emph{human-centered representativeness} are not identical. A convincing conversational persona may still fail to preserve cultural nuance, user values, or normative intent. Thus, the future challenge is not only to simulate \emph{human-like} behavior, but to simulate \emph{human-centered} behavior grounded in real social and ethical constraints.

\section{Comparison of Traditional and Human-Centered Agents}

Table~\ref{tab:comparison} summarizes the central contrast developed in this survey.

\begin{table*}[t]
\centering
\small
\begin{tabular}{p{2.9cm}p{5.2cm}p{6.1cm}}
\toprule
\textbf{Dimension} & \textbf{Traditional / Task-Centric Agents} & \textbf{Human-Centered Agents} \\
\midrule
Primary Objective & Task completion, optimization, prediction & Context-sensitive representation and action on behalf of humans \\
Decision-Making & Utility-driven or heuristic task planning & Bounded rationality, reflective reasoning, context-aware judgment \\
Reasoning Model & Logical or statistical inference & Hybrid cognition: common sense, dual-process elements, mental models \\
Language Use & Fluent, often generic, instruction-following & Pragmatic, audience-aware, culturally adaptive communication \\
Cultural Awareness & Minimal or implicit & Explicit modeling of norms, regional context, and sociolinguistic variation \\
Values and Beliefs & Often absent or collapsed into generic safety alignment & Preference-aware, pluralistic, belief-sensitive, norm-constrained decision-making \\
Human Interaction & User as requester or supervisor & Human as teammate, stakeholder, or represented principal \\
Explainability & Optional add-on & Core requirement for trust, coordination, and accountability \\
Coordination in MAS & Task-level communication and synchronization & Socially grounded coordination including roles, trust, negotiation, and norm awareness \\
User Modeling & Static profiles or none & Dynamic representations of preferences, beliefs, history, and social context \\
Evaluation & Accuracy, efficiency, benchmark performance & Human-centered success: appropriateness, trustworthiness, cultural fit, collaboration quality \\
\bottomrule
\end{tabular}
\caption{Conceptual comparison between traditional task-centric agents and the human-centered agent perspective advocated in this survey.}
\label{tab:comparison}
\end{table*}

\section{Illustrative Application Scenario}

To make the discussion concrete, consider a multicultural healthcare support environment in which multiple AI agents assist clinicians, patients, and administrators.

A task-centric system might perform well on information retrieval, appointment scheduling, or treatment recommendation. Yet such a system could still fail in important ways: it may communicate too directly in one cultural setting and too vaguely in another; it may ignore patient beliefs about treatment, family involvement, or authority; it may explain decisions in a technically correct but socially ineffective manner; or it may coordinate with other agents in a way that satisfies a global optimization objective while violating local value expectations.

A human-centered system would need to:
\begin{itemize}[leftmargin=*]
    \item represent patient preferences and culturally shaped expectations,
    \item adapt language and explanation style to audience and context,
    \item communicate uncertainty to clinicians transparently,
    \item preserve confidentiality and ethical constraints,
    \item coordinate among agents while maintaining shared mental models and trust.
\end{itemize}

This example illustrates why human-centeredness is not an optional layer of personalization. In many real domains, it is integral to competence.

\section{Open Challenges and Future Research Directions}

The literature points toward several urgent research directions.

\subsection{Unified Cognitive-Social Agent Architectures}

Current systems typically specialize in one dimension: cognitive modeling, cultural adaptation, alignment, or collaboration. Future work should aim at unified architectures in which memory, planning, social context, value constraints, and communication policies are modeled jointly rather than bolted together.

\subsection{Representing Cultural Context Without Stereotyping}

Cultural adaptation is necessary, but simplistic cultural labeling can lead to overgeneralization or stereotyping. Research is needed on probabilistic, interaction-sensitive, and self-updating models of culture that treat users as individuals embedded in social contexts rather than as fixed demographic templates.

\subsection{Pluralistic and Dynamic Value Alignment}

Alignment research increasingly recognizes value pluralism, but scalable implementations remain difficult. Future agents must balance universal safety requirements with individual and community-specific preference adaptation, including mechanisms for revising value models over time.

\subsection{Robust Social Reasoning}

Theory of Mind-like capabilities, negotiation strategies, and shared mental model tracking are promising, but current evaluations remain mostly benchmark-centric. Future work needs interactive, ecologically valid tests of social reasoning in long-horizon human-agent and agent-agent collaboration.

\subsection{Trust-Aware Multi-Agent Coordination}

Trust is not just a human-agent issue. In multi-agent societies, trust calibration, reputation, and accountability may affect whether collaborative structures remain aligned and stable. This suggests the need for MAS frameworks that explicitly model trust, role legitimacy, and social responsibility.

\subsection{Evaluation Beyond Task Accuracy}

Human-centered agents require new evaluation criteria. Important metrics likely include:
\begin{itemize}[leftmargin=*]
    \item cultural appropriateness,
    \item value conformity under pluralistic settings,
    \item explanation usefulness,
    \item collaboration quality,
    \item persona consistency,
    \item robustness of social reasoning,
    \item long-term trustworthiness.
\end{itemize}

Benchmarks like CDEval are a useful start \citep{wang2024cdeval}, but the field still lacks comprehensive evaluation frameworks that cover all major human-centered dimensions.

\subsection{Ethics, Governance, and Societal Oversight}

Finally, human-centered agent research cannot remain purely technical. Agents acting in socially embedded roles raise questions of accountability, consent, representation, fairness, governance, and institutional legitimacy. Building such systems responsibly requires interdisciplinary collaboration across AI, HCI, psychology, sociology, linguistics, and ethics.

\section{Discussion}

A coherent pattern emerges across the literature surveyed here. Advances in LLMs and autonomous agents have made it possible to build systems that appear increasingly intelligent in traditional computational terms. They can reason, converse, plan, retrieve information, invoke tools, and collaborate in distributed workflows. However, this progress has exposed a deeper limitation: intelligence alone does not yield effective participation in human environments.

Human environments are not merely informational they are social, cultural, and normative. People do not decide only by calculating utility; they reason under constraints, emotion, uncertainty, and social expectation. They do not communicate only through literal semantics; they speak through context, audience, and culture. They do not collaborate simply by dividing labor; they cooperate through trust, mutual modeling, explanation, and negotiated norms. And when they act on behalf of others, they must preserve values, beliefs, and responsibilities.

The literature has begun to address each of these dimensions:
\begin{itemize}[leftmargin=*]
    \item cognitive science provides theories of bounded rationality, dual-process reasoning, common sense, and theory of mind;
    \item cultural and sociolinguistic research clarifies how communication and behavior vary across contexts;
    \item alignment and preference learning provide tools for modeling values and human intentions;
    \item HCI and collaborative AI emphasize trust, explainability, and teaming;
    \item MAS research provides infrastructures for distributed interaction;
    \item cognitive architectures, ABM, persona modeling, and generative agents offer implementation pathways.
\end{itemize}

Yet the key problem remains integration. Existing systems typically combine strong task competence with shallow representations of the human dimensions in which they operate. As a result, they can be competent but socially brittle, fluent but culturally narrow, aligned in the aggregate but not in the particular, collaborative in protocol but not in understanding.

This survey therefore positions the next frontier of AI agents not simply as greater autonomy, but as \emph{human-centered autonomy}: autonomy constrained and informed by cognition, culture, values, and cooperation.

\section{Conclusion}

This survey has argued that the evolution of AI agents from rule-based systems to LLM-based autonomous and multi-agent architectures---has created an opportunity and a necessity. The opportunity lies in building systems that can participate meaningfully in complex human environments. The necessity lies in confronting the fact that such participation requires more than language generation and task optimization.

We reviewed six major literatures: the evolution of intelligent agents, human cognition and decision-making, language and culture, human values and belief systems, human-agent collaboration, and multi-agent coordination with computational models of human behavior. Across these areas, the same conclusion recurs: the most important properties of human-centered interaction are still weakly represented in current agents.

The central research gap can therefore be stated as follows:

\begin{quote}
\textbf{Existing AI and multi-agent systems primarily model intelligence as prediction, optimization, and task completion. However, agents intended to act on behalf of humans in cooperative environments require deeper computational representations of human cognition, cultural context, value systems, and social behavior. Despite substantial advances in individual areas such as cognitive modeling, cultural intelligence, and AI alignment, there remains no widely adopted framework that integrates these dimensions into autonomous agents capable of authentic human-centered collaboration.}
\end{quote}

Addressing this gap is likely to define a major part of the next stage of agent research. The future of multi-agent systems is not simply more agents or more autonomy. It is agents that can reason with, communicate like, and collaborate on behalf of humans in ways that are contextually appropriate, culturally aware, value-aligned, and socially trustworthy.

\bibliography{custom}

\end{document}